\documentclass[11pt,a4paper]{article}
\usepackage[hyperref]{emnlp-ijcnlp-2019}
\usepackage{cleveref}
\usepackage{times}
\usepackage{latexsym}
\usepackage{graphicx}
\usepackage{multirow}
\usepackage{pgfplots}
\usepackage{url}
\usepackage{balance}

\aclfinalcopy 

\title{A Study of BERT for Non-Factoid Question-Answering under Passage Length Constraints}

\author{Yosi Mass, Haggai Roitman, Shai Erera, Or Rivlin, Bar Weiner and David Konopnicki \\
  IBM Research AI\\
  Haifa University, Mount Carmel, Haifa, HA 31905, Israel \\
  {\tt \{yosimass,haggai,shaie,or.rivlin,barw,davidko\}}@il.ibm.com}

\date{}

\begin{document}
\maketitle
\begin{abstract}
We study the use of BERT for non-factoid question-answering, focusing on the passage re-ranking task under varying passage lengths. To this end, we explore the fine-tuning of BERT in different learning-to-rank setups, comprising both point-wise and pair-wise methods, resulting in substantial improvements over the state-of-the-art.
We then analyze the effectiveness of BERT for different passage lengths and suggest how to cope with large passages. 
\end{abstract}

\section{Introduction}
\label{sec:introduction}


Within a question-answering (QA) setting, the passage-retrieval task, retrieves a (relatively) short text fragment (passage)
that provides a focused answer to a given question.
As opposed to the factoid-QA task, which requires a succeeding step of answer extraction, the non-factoid QA task aims at retrieving passages as the answers.

In this work, we focus on the passage retrieval task and more specifically, on the passage re-ranking task, i.e., given an initial ranked-list of passages, retrieved by some basic passage-retrieval method (e.g., BM25~\cite{psgbm25}), our goal is to re-rank the passages in the list so as to position the most relevant ones higher. Similar to the latest works on this task, our approach is based on Deep-Learning (DL).
Existing DL methods commonly train their networks directly on the labeled datasets~\cite{hybrid:2018, wikiPassageQA:2018,mitra2019duet,tan2015lstmbased}. While achieving better results than more ``traditional'' methods, such methods still suffer from the relatively small labeled data available for training.

Trying to overcome such data limitations, more recent works have utilized the pre-trained Bidirectional Encoder Representations from Transformers (BERT) language representation models~\cite{BERT:2018}. 
Each BERT model was trained on large corpora of Wikipedia and news data, 
consisting of a large number of layers (12/24 in the base/large model) based on the Transformer~\cite{Transformer:2017} architecture.
When trained on pairs of sentences, BERT allows to capture high-order interactions between a given pair of sentences. Such pre-trained models can be then fine-tuned for a specific downstream task with a relatively small amount of labeled data. Previously, such end-to-end networks have been shown to be highly effective in several NLP tasks, and very recently also for passage-retrieval~\cite{multiTaskBert:2019, passageRerankingBert:2019}.

While utilizing BERT for passage re-ranking is promising, there are still several technical limitations that are imposed by BERT's basic architecture. First, due to the large number of layers of BERT, it should be run on a GPU, enforcing a maximum of 512 input tokens 
per pair.
Among these tokens are the special \textsf{[CLS]} token and two \textsf{[SEP]} tokens that BERT adds to each pair. Moreover, BERT's tokenizer breaks tokens not in its dictionary into sub-tokens, which further limits the total number of tokens that can be passed for each pair.

The primary goal of our work is to study how BERT can be best utilized for the passage re-ranking task, within the non-factoid QA setting, in spite of its input length limitations. To this end, we report on an extensive empirical analysis with different passage lengths. We show that, too short passages may suffer from a lack of content, while too large passages may generate noisy representations. We further report on experiments that try to breakdown passages into smaller chunks and combine their representations into a passage-level representation.
\section{Related work}
\label{sec:related}

Following the success in applying BERT~\cite{BERT:2018} for multiple NLP tasks, some very recent works have applied it to
the factoid QA~\cite{BERTserini} task, as well as to the non-factoid QA task~\cite{multiTaskBert:2019,passageRerankingBert:2019,qiao2019understanding}. 
In this work we focus on the non-factoid QA task.

Similar to the MRPC paraphrase task in the original BERT work, the authors of~\cite{passageRerankingBert:2019} have fine-tuned BERT for passage re-ranking by casting it as a binary classification problem. To this end, correct passages were used as positive examples and the rest as negative examples. At run time, the classification score of passages was used for ranking. In~\cite{multiTaskBert:2019}, the authors further fine-tuned BERT with a variation of the triplet-loss for ranking passages within a factoid QA task.

In this work we study the behaviour of BERT with respect to passage length properties. 
A couple of recent works have also investigated BERT's behaviour for retrieval tasks. In~\cite{qiao2019understanding} an analysis of BERT's attention allocation between query-document tokens in its Transformer layers was performed, further examining its difference from soft match patterns learned by a traditional learning-to-rank (LTR) neural model. The work by~\cite{padigela2019investigating} has compared BERT's performance to that obtained by a BM25 model with respect to various query properties (e.g., query-length and term frequencies).


\section{Passage Re-ranking with BERT}
\label{sec:triplets}

Here, we shortly discuss several options for fine-tuning BERT for the passage re-ranking task.
To this end, we exploit a given pre-trained BERT model\footnote{ https://github.com/google-research/bert} for assigning an effective representation $\texttt{BERT}(s_1,s_2)$ for any given pair of sentences $(s_1,s_2)$. As the representation itself, we consider it to be that of the \textsf{[CLS]} token~\cite{BERT:2018} which serves as the aggregated pair representation.

We now assume the availability of some labeled examples of queries, where each query ($q$) is potentially associated with both positive ($p^{+}$) and negative ($p^{-}$) passages.
As a first step, we create training examples, consisting of (positive and negative) query-passage pairs for fine-tuning BERT.
In our study we explore two main learning-to-rank~\cite{Li2011ASI} (LTR) strategies with BERT, namely \textbf{point-wise} and \textbf{pair-wise}.

\paragraph{Point-wise}
Within a point-wise LTR approach, each passage is scored independently of other passages~\cite{Li2011ASI}. To this end, we use the MRPC (Microsoft Research Paraphrase Corpus) classifier as described in~\cite{BERT:2018}. The classifier has two labels, $\textsf{0}$ - for negative examples (i.e., ($q,p^{-}$) pairs) and $\textsf{1}$ - for positive examples (i.e., ($q,p^{+}$) pairs). Training the LTR model is simply implemented by minimizing the cross-entropy loss. Once trained, the classifier's confidence is used as the score. We refer to this ranking method as \texttt{BERT[PW]}.

\paragraph{Pair-wise}
Compared to the point-wise LTR approach, a pair-wise LTR approach is trained to learn a preference (ordering) among pairs of passages~\cite{Li2011ASI}.
In this work we implement two such methods.  The first method, termed \texttt{BERTlets}, uses a \textit{Triplet-Network}~\cite{Triplet:2018} as follows.
Let $v$ be a vector, having the same dimension as the obtained BERT representation\footnote{$768$/$1024$ for the base/large BERT model.}. We define the score of a passage $p$ for query $q$ as:
\vspace{-0.1in}
\begin{equation}\label{eq:score}
  score(q;p) = \texttt{BERT}(q,p) \cdot v,
\end{equation}

where \texttt{BERT}(q,p) is the BERT pooled representation of the CLS token of the last layer.

Given a triplet $t = (q,p^{+},p^{-})$, 
let $s^{+} = score(q;p^{+})$ and $s^{-} = score(q;p^{-}))$ be the scores assigned to the positive and the negative passage samples for a given query $q$, respectively.
Our goal is to tune $v$, such that 
$(q,p^{+}) \prec (q,p^{-})$ (or $s^{+} > s^{-}$). To achieve that, we train the network with a Hinge loss~\cite{hinge}:
\begin{equation}\label{eq:loss}
   loss_t = \max(m - (\hat{s}^{+} -\hat{s}^{-}), 0),
\end{equation}

where $\hat{s}^{+} = \frac {\exp{s^{+}}}{\exp{s^{+}}  + \exp{s^{-}}}$, $\hat{s}^{-} = \frac {\exp{s^{-}}}{\exp{s^{+}}  + \exp{s^{-}}}$ and $m$ is the margin hyperparameter (e.g., 0.2). 

The second pair-wise method, termed \texttt{BERT[CE]}, is inspired by~\cite{multiTaskBert:2019} and serves as an alternative to \texttt{BERTlets}. This method uses a similar technique as \texttt{BERTlets}, but instead of triplets, it is trained by comparing a given positive pair $(q,p^{+})$ with any given negative pair $(q,p^{-})$. It also learns a vector $v$ and assigns a similar $score(q,p)$ as in Eq.~\ref{eq:score}. Yet, instead of the Hinge loss, its loss is the negative log-likelihood of the positive example (i.e., $-log (\hat{s}^{+})$). 



\section{Experimentation Analysis}
The primary goal of our analysis is to evaluate the behavior of the three proposed BERT-based LTR methods under various passage-length and segmentation settings.
Our implementation of the various LTR models is based on the TensorFlow\footnote{https://www.tensorflow.org/} version of BERT\footnote{ https://github.com/google-research/bert}. 
In all the experiments we used the pre-trained BERT-Base, Uncased model (12-layer, 768-hidden, 12-heads, 110M parameters).
 
We start with a short description of our datasets and setup, followed by the empirical results of the analysis.
\label{sec:experiments}

\subsection{Datasets}
\label{sec:datasets}
We used three different datasets that were previously used for the non-factoid QA task~\cite{hybrid:2018,wikiPassageQA:2018}. These datasets are next shortly described while their statistics are summarized in Table~\ref{tab:datasets}.

\noindent \textbf{nfL6}\footnote{https://ciir.cs.umass.edu/downloads/nfL6}~\cite{hybrid:2018} consists of 87,361 Yahoo's non-factoid questions. Each query has a single correct passage.

\noindent \textbf{WebAP}\footnote{https://ciir.cs.umass.edu/downloads/WebAP/}~\cite{hybrid:2018} consists of 82\footnote{Two queries have no relevant answers so are omitted from our experiments.} queries with labeled passages from the TREC GOV2 web collection. Queries in this dataset are more open ended than the former dataset, and can have a variety of relevant passages. 
Labeling was done by first retrieving the top-50 web documents for each query and then marking contiguous sequences by a five graded-relevance scores (4 - Perfect, 3 - Excel, 2- Good, 1 - fair, 0 - None). Similar to~\cite{hybrid:2018}, we took the relevant sequences (i.e., relevance $>0$) as positive passages and divided the irrelevant sequences to negative ones. Since there was not enough details in~\cite{hybrid:2018} on how the latter was done, we assumed passage lengths are normally distributed with mean and standard deviation estimated using the relevant passages.

\noindent \textbf{WikiPassageQA}\footnote{https://ciir.cs.umass.edu/downloads/wikipassageqa/}~\cite{wikiPassageQA:2018} contains 4,165 non-factoid queries with passages extracted from Wikipedia articles. Here, it is assumed that for each query, there is a single Wikipedia article that contains relevant passages, which are manually marked as contiguous sequences. Unlike WebAP, each article is divided to fixed passages of $6$ sentences each, where a passage is labeled relevant if it has at least $15\%$ overlap with a marked relevant sequence.

\begin{table}[tbh]
	\center
    \small
		\caption{Datasets}
        \setlength\tabcolsep{2.0pt}
		\begin{tabular}{l|c|c|c|c|c|}
		    \cline{2-6}
			 & \textbf{queries} &\textbf{psgs} & \textbf{min} &
			    \textbf{max} & \textbf{avg}   \\
			\hline
			\multicolumn{1}{|l|}{nfL6} &
			87,362 &  87,362 & 2 & 809  & 42.4  \\
			\hline
			\multicolumn{1}{|l|}{WebAP} &
			80 & 489,042 & 0 & 10,851 & 74.5 \\
             \hline
			\multicolumn{1}{|l|}{WikiPassageQA} &
			4,165 & 50,612 & 10 & 1,332 & 134.2\\
			\hline
		\end{tabular}
	  \label{tab:datasets}
\end{table}

\subsection{Setup}
\label{sec:train}
Similar to~\cite{hybrid:2018}, for each query, we 
retrieved the top-$k$ BM25 results in the nfL6 and WebAP datasets ($k=10$ for nfL6 and $k=100$ for WebAP). To this end, we indexed each passage as a document in an ElasticSearch\footnote{https://www.elastic.co} index. For WikiPassageQA, we simply took all passages from the query's Wikipedia article. 

For fine-tuning the three 
BERT models, we used the retrieved top-$k$ passages of each query $q$, and created their corresponding triplets set as follows. 

Given a query $q$ with $m$ positive examples and $n$ negative examples (usually $m << n$), each positive passage participates in different $n / m$ triplets.
To cope with a reasonable number of triplets, we down-sampled nfL6 (which has a large number of queries) to have at most two negative examples for each positive one.  For WikiPassageQA we down-sampled to five negative examples per each positive one. Since WebAP has a small number of queries, we used all the negative examples.

WikiPassageQA is already divided into $3,332$ train, $417$ development and $416$ test queries. For the other two datasets, we used a 5-fold cross validation. Similar to~\cite{BERT:2018}, we used a fixed number of three train epochs, hence we omitted the development set.

To evaluate the performance of the various ranking methods, we re-ranked the top-$k$ passages returned by the BM25 baseline ranker. Similar to~\cite{hybrid:2018}, for the nfL6 and WebAP  datasets, we added a correct passage at position $k$ ($k=10$ for nfL6 and $k=100$ for WebAP) if it was not retrieved within the BM25 top-$k$ passages.

We used standard Information Retrieval (IR) metrics: Precision@1 (P@1), Mean Average Precision (MAP) and Mean Reciprocal Rank (MRR).

Training was done using the default BERT setup~\cite{BERT:2018}, with Adam optimizer, learning rate $2e$-$5$ over three epochs.
All experiments were run on a Tesla K40m GPU, with a memory clock-rate of 0.745Ghz and 12Gb memory.

\subsection{Results}
\label{sec:results}

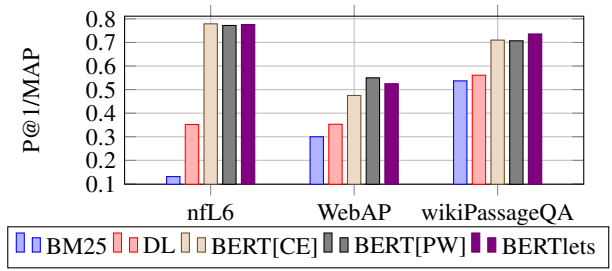
\begin{figure}[tbh]
\small
\begin{tikzpicture}
\begin{axis}[
    ybar,
		grid=major,
    ytick  ={0.1,0.2,0.3,0.4,0.5,0.6,0.7,0.8,0.9,1.0},
    yticklabels={0.1,0.2,0.3,0.4,0.5,0.6,0.7,0.8,0.9,1.0},
		width=3in,
		height=1.5in,
		enlargelimits=0.3,
		enlarge y limits=0.05,
    legend style={at={(0.4,-0.25)},
      anchor=north ,legend columns=-1},
    ylabel={P@1/MAP},
    symbolic x coords={nfL6,WebAP,wikiPassageQA},
    xtick=data,
		every node near coord/.append style={font=\tiny},
		bar width=5pt
    ]
	\addplot coordinates {(nfL6,.131) (WebAP,.3) (wikiPassageQA,0.537)};
	\addplot coordinates {(nfL6,.352) (WebAP,.353) (wikiPassageQA,0.561)};
	\addplot coordinates {(nfL6,.779) (WebAP,.475) (wikiPassageQA,0.71)};
	\addplot coordinates {(nfL6,.772) (WebAP,.55) (wikiPassageQA,0.707)};
	\addplot coordinates {(nfL6,.776) (WebAP,0.525) (wikiPassageQA,0.736)};
	\legend{BM25, DL, BERT[CE], BERT[PW], BERTlets}
\end{axis}
\end{tikzpicture}
\caption{Comparison between methods (P@1 and MAP). SeqLen=$256$}
\label{fig:all-map}
\end{figure}

We compared the three studied BERT-based LTR methods (Section~\ref{sec:triplets}) to two baselines, namely BM25 and the previously best known deep-learning (DL) method. Similar to the reported DL methods, we use P@1 for the first two datasets (nfL6 and webAP)~\cite{hybrid:2018} and MAP for the third one (WikiPassageQA)
~\cite{,wikiPassageQA:2018}. Figure~\ref{fig:all-map} depicts the results obtained by each ranker.  The three BERT methods use 256 tokens for each pair (SeqLen in the BERT jargon), which was found to be the optimal size, as we discuss in Section~\ref{sec:block_size} below.

We can observe that the three BERT methods significantly outperform the two baselines. The improvements range from $31\%$ improvement on MAP in WikiPassageQA to $120\%$ on P@1 in nfL6 over the best DL baseline.
Among the three BERT methods, the point-wise
(\texttt{BERT[PW]}) performed better on webAP (with P@1=$0.55$, compared to $0.53$ by the pair-wise (\texttt{BERTlets})), while \texttt{BERTlet} performed better on WikiPassageQA (with MAP=$0.74$ compared to $0.71$ by \texttt{BERT[PW]}).  
Among the two pair-wise BERT variants, the \texttt{BERTlets} method performs better. 
The experiments in the subsequent sections are based on \texttt{BERTlets}.

\subsubsection{Effect of Passage Length}
\label{sec:block_size}

\begin{figure}[tbh]
\begin{tikzpicture}
	\begin{axis}[
		width=3in,
		height=1.5in,
		symbolic x coords={n=64,n=128,n=256,n=384},
		ylabel=P@1/MAP,
		ymin=0.4,
		legend style={font=\small,at={(0.4,-0.22)},
									anchor=north,legend columns=-1},
		xtick=data,
    nodes near coords align={vertical},
		axis y line*=left
	]
	\addplot+[mark=square,very thick] coordinates {
		(n=64,0.735)(n=128,0.779)(n=256,0.776)(n=384,0.779)
	};
	\addlegendentry{nfL6}
		
	\addplot+[mark=o,very thick] coordinates {
		(n=64,0.463)(n=128,0.538)(n=256,0.525)(n=384,0.525)
	};
	\addlegendentry{WebAP}

	\addplot+[mark=triangle,very thick] coordinates {
		(n=64,0.5)(n=128,0.67)(n=256,0.736)(n=384,0.714)
	};
	\addlegendentry{wikiPassageQA}

	\end{axis}
\end{tikzpicture}
\caption{Effect of SeqLen (n) on BERTlets}
\label{fig:passage_size_map}
\end{figure}
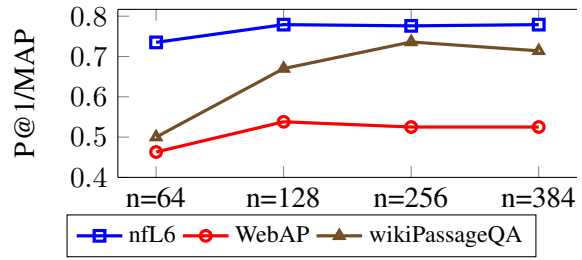

As was mentioned above, BERT imposes a limit of maximum $512$ tokens
\footnote{BERT ignores input tokens after the first SeqLen tokens} (SeqLen) in each pair when using a GPU. Figure~\ref{fig:passage_size_map} 
depicts the 
\texttt{BERTlets} results for varying SeqLen. We observe that, the best P@1 and MAP are achieved for $256$ tokens.  Lower values ($64$ and $128$) actually hurt the performance while increasing to $384$ does not improve and even hurts it mainly on WikiPassageQA.

The effect of the small SeqLen is mainly manifested on the WikiPassageQA which has the largest average passage-length as shown in Table~\ref{tab:datasets}. This is actually expected, since short passages may lack enough content, while too long ones may include additional superfluous content (``noise'') to allow to fully capture such passages semantics through the BERT representation.


\subsubsection{Effect of passage segmentation}
\label{sec:sentences}

To overcome the limitations of BERT on passage-length, we further segmented each passage into equal number of chunks. Given a query $q$ and passage $p$, we derived $\texttt{BERT}(q,c)$, for chunks $c \in p$,  and combined their representation through an attention mechanism~\cite{Attention:2016}.  We tried to break to two and three chunks and pass them with SeqLen$=128$, using an attention size of $192$. The results are summarized in Table~\ref{tab:sentences}.

\begin{table}[h]
	\center
    \small
		\setlength\tabcolsep{1.8pt}
		\caption{Effect of breaking paragraphs to chunks of SeqLen$=128$ (with attention)}
		\begin{tabular}{l|c|c||c|c||c|c|}
			\cline{2-7}
			\multirow{2}{*}{} &
				\multicolumn{2}{c||}{nfL6} & 																					\multicolumn{2}{c||}{WebAP} &
				\multicolumn{2}{c|}{WikiPassageQA} \\
			\cline{2-7}
 					& P@1 & MRR & P@1 & MRR & MAP & MRR \\ \hline
					\multicolumn{1}{|l|}{BERTlets (2*128)} & .744 & .835 & .513  & .662 & .715 & .778  \\ \hline
					\multicolumn{1}{|l|}{BERTlets (3*128)} & .748 & .839 & .488 & .634 & .715  & .791 \\ \hline
					\multicolumn{1}{|l|}{BERTlets (256)} & .776 & .857 & .525 & .683 & .736 & .810 \\ \hline
		\end{tabular}
    \label{tab:sentences}
\end{table}

As we can observe, using passage segmentation results in a slightly lower performance. 
However, given the BERT constraints on passage length, this segmentation can be used to handle large passages above the BERT limitation of $512$ tokens. We further tried to combine the chunks representations using \textit{Max-Pooling} but it gave similar results to the attention method.

\section{Conclusions}
\label{sec:conclusions}
We have studied the utilization of BERT for the task of passage re-ranking in light of its strict limitation on passage length. Using three BERT-based LTR methods, we conducted experiments on three datasets, with different passage lengths and segmentation configurations.

Our findings are that using BERT representations for mid-sized (SeqLen=256 tokens) of (query, passage) pairs give the best results. We have further shown that, by breaking passages to smaller chunks and aggregating their (query, chunk) representations, we can re-rank larger passages, with only a moderate degradation in quality. Finally, we found out that point-wise and pair-wise LTR achieved similar results on the three datasets. 

\balance

\bibliography{BERTlets}
\bibliographystyle{acl_natbib}

\end{document}